\documentstyle[12pt]{article}
\begin{document}

\title{Critical behaviour of a surface reaction model with
infinitely many absorbing states}

\author{Iwan Jensen \\ Department of Mathematics,
The University of Melbourne, \\ Parkville, Victoria 3052,
Australia. \\ e-mail: iwan@mundoe.maths.mu.oz.au }
 
\maketitle

\begin{abstract} 
In a recent letter [J. Phys. A {\bf 26}, L801 (1993)],
Yaldram {\em et al} studied the critical behaviour of a simple
lattice gas model of the CO-NO catalytic reaction. The model
exhibits a second order nonequilibrium phase transition 
from an active state into one out of 
infinitely many absorbing states. Estimates for the critical 
exponent $\beta$ suggested that the model belongs to a new 
universality class. The results reported in this article
contradict this notion, as estimates for various
critical exponents show that the model belongs
to the universality class of directed percolation. 
\end{abstract} 


\newpage

Nonequilibrium phase transitions occur in many models studied
in physics, chemistry, biology or even sociology. A special
group of models, that have attracted a great deal of interest
in recent years, exhibit a continuous transition into an absorbing
state. The best known examples are probably 
directed percolation (DP) \cite{dpbook}--\cite{kinzel}, 
Reggeon field theory \cite{gribov,brower}, 
the contact process \cite{harris}--\cite{durrett}, 
and Schl\"{o}gl's first and second models
\cite{schlogl}--\cite{pgrg}. Extensive studies of 
these and many other models \cite{zgb}--\cite{zhuo} with a 
{\em unique} absorbing state have revealed that 
they belong to the same universality class. This 
provides firm support for the conjecture that continuous transitions 
into a unique absorbing state generically belong to the DP class 
\cite{janssen81,pgrg}.

For models with multiple absorbing states the situation is not
so simple. Some studies of two-dimensional surface reaction models
yield critical exponents different from those of directed 
percolation in (2+1)-dimensions
\cite{dimertrimer,dimerdimer}. However, the pair contact process
(PCP) and dimer reaction (DR) model (in one dimension) clearly
belongs to the DP universality class \cite{pcpprl,pcppre}, at least
as far as the {\em static} critical behaviour is concerned.
In all of these models the number of absorbing configurations grows
exponentially with system size. However, all of the absorbing
configurations are characterized by the vanishing of a unique
quantity, e.g., the number of particle pairs in the PCP or 
in other cases the \cite{dimertrimer,dimerdimer}
number of nearest neighbor vacancy pairs.

Recently, Yaldram {\em et al.} \cite{cono}, studied the critical
behaviour of a simple lattice model of the CO-NO catalytic
reaction in which CO + NO $\rightarrow$ CO$_{2}$ + $\frac{1}{2}$N$_{2}$. 
Schematically the reaction steps are given as:

\begin{eqnarray}
     CO^{g} + * & \rightarrow & CO^{a},  \\ 
     NO^{g} + 2* & \rightarrow & O^{a} + N^{a}, \\
     CO^{a} + O^{a} & \rightarrow & CO_{2}^{g} + 2*,  \\
     N^{a} + N^{a} & \rightarrow & N_{2}^{g} + 2*, 
\end{eqnarray}

where the superscript {\em g (a)} refers to a molecule in the
the gas fase (adsorbed on the surface) and $*$ marks an empty site.
The catalytic surface is modelled by a two-dimensional triangular
lattice. The rules of the computer algorithm are quite simple,
with probability $p$ a CO molecule is adsorbed on an empty site
and with probability $1-p$ NO adsorption is attempted. Since NO
dissociates upon adsorption it requires a nearest neighbour pair
of empty sites. In the simulations this is done by first chosing
an empty site at random and then chosing one of the six nearest 
neighbours randomly, if the neighbour is empty O is placed on the
original site and N on its neighbour. After each adsorption the
nearest neighbours are checked (in random order) and CO+O reacts
to form CO$_{2}$ which leaves the surface at once, likewise
N+N forms N$_{2}$ which desorbs immediately. It is thus obvious 
that any state without empty sites is absorbing. All processes
depend on the presence of empty sites so an efficient algorithm uses
a list of these. After each attempted adsorption the time variable
is incremented by $1/N_{e}$, where $N_{e}$ is the number of empty
sites prior to the attempt, thus making each time step equal to
(on the average) one attempted update per lattice site. The algorithm
outlined above differs from that used by Yaldram {\em et al.} in
one aspect, when NO adsorbs they choose a {\em pair} of empty sites
at random, whereas I choose one empty sites and a nearest neighbor
and only adsorb NO if the nearest neighbor is empty. This makes
NO adsorption less likely in my algoritm. However, one would expect
this merely to lead to a change in the location of the phase 
transitions {\em not} to a change in the critical behaviour.
Computer simulations
by Yaldram {\em et al.} \cite{cono} shows that when $p < p_{1}$ 
the system always enters an absorbing state in which the lattice 
is covered by a mixture of O and N (but off course without any 
nearest neigbour pairs of N). Note that the symmetry of the lattice
prevents a CO from being surrounded by N, as some of these N would
have to be nearest neighbours and thus react. 
The number of absorbing configurations grows
exponentially with system size. Note also that an absorbing 
configuration, though not unique, is characterized by the vanishing
of the number of empty sites. At $p_{1}$ the model exhibits a 
{\em continuous} phase transition into an active state in which 
the catalytic process can prodeed indefinetely. Finally when
$p$ exceeds a second critical value $p_{2}$ the model exhibits
a {\em discontinuous} phase transition into a CO and N covered state.
The phase diagram of the CO-NO reaction model is thus very similar
to that found in various similar catalytic model 
\cite{zgb,dimertrimer,dimerdimer}. Near the critical point
$p_{1}$ one would expect the concentrations $\rho_{X}$ of 
various lattice sites $X$ ($X$ = O, N, CO, or an empty site) to 
follow simple power laws,

\begin{equation}
    \rho_{X} - \rho_{X}^{sat} \propto (p-p_{1})^{\beta_{X}},
\end{equation}

where $\rho_{X}^{sat}$ is the saturation concentration. Note
that the saturation concentration for empty sites and CO is zero, 
whereas it is non-zero for O and N. Yaldram {\em et al.} \cite{cono}
found that $p_{1} = 0.185(2)$, where the figure in parenthesis is 
the uncertainty in the last digit, and $\beta_{X} = 0.20-0.22$.
The estimates for $\beta_{X}$ are much smaller that the value
$\beta = 0.592(10)$, obtained using 
the scaling relation $\beta = \delta\nu_{\parallel}$ \cite{torre} 
with $\delta=0.460(6)$ and $\nu_{\parallel}=1.286(5)$
\cite{pgdp}, for directed percolation in
(2+1)-dimensions. This could indicate that the CO-NO model 
belongs to a new universality class. However, the uncertainty in
the estimate for $p_{1}$ is quite large, especially considering
that the $\beta$ estimates are obtained using values of $p-p_{1}$ 
between 0.01 and 0.001, which overlaps the error estimate for 
$p_{1}$. Moreover, the lattice sizes ($40\times 40$) used in the 
simulations are very small. Actually 
for such small lattice sizes one would expect finite-size effects 
to be quite prominent. All in all I think there is ample reason to 
doubt the validity of the exponent estimates obtained by 
Yaldram {\em et al.}.

In this article I report the results of extensive simulations
of the CO-NO model using time-dependent simulations
and finite-size scaling. The general idea of time-dependent 
simulations is to start from a configuration which is very close to
the absorbing state, and then follow the ``average" time evolution
of this configuration by simulating a large ensemble of independent
realisations. This method is straight forward and very 
successful for models with a unique absorbing state 
\cite{torre,bbc,tripann,tripcrea,pgdp}. For models with
multiple absorbing state the situation is more intricate, as a
recent study \cite{pcppre} revealed that the {\em dynamic} critical
exponents predicted via time-dependent simulations depend upon 
the choice of initial configuration. However, two important facts
emerged from this study, first of all the predictions for the
location of the {\em critical point} was allways correct, and
secondly if one uses an initial configuration reminiscent of a
{\em typical} absorbing configuration the predictions for the
dynamical critical exponents coincide with those expected from the 
static critical behaviour. A recent more thorough study by Mendes
{\em et al.} \cite{mulscal} have confirmed this picture and led to 
a generalized scaling ansatz for models with multiple absorbing 
states. In this study I generate the initial configuration by 
simulating the CO-NO model on a 128$\times$128 lattice (with 
periodic boundary conditions) at the value of $p$ under 
investigation until it enters an absorbing state. An off-set 
$(x,y)$ is then chosen randomly on this lattice. Hereafter the 
configuration is mapped cyclically onto a larger (512$\times$512) 
lattice such that $(x,y)$ is at the origin of the larger lattice.
The particle at position $(i,j)$ on the large lattice is the 
same as the particle at position $(i+x \bmod 128,j+y \bmod 128)$ on 
the small lattice. Hereafter a pair of empty sites is placed at 
the origin. The size of the large lattice ensures that the cluster
of empty sites grown from the seed at the origin never reaches the
boundaries of the lattice. 
We thus start in a configuration close to an absorbing 
state (just two sites are open) and it should be close to a typical
absorbing state of the infinite system. For each such configuration
I simulated 5000 independent samples and typically 50--100 independent
configurations for a total of 250-500,000 samples. Each run had
a maximal duration of 2000 time steps, but most samples enters an
absorbing state before this limit is reached. As usual in this type
of simulation I measured the survival probability $P(t)$, the
average number of empty sites $\bar{n}(t)$, and the average mean
square distance of spreading $\bar{R}^{2}(t)$ from the origin.
Notice that $\bar{n}(t)$ is averaged over all runs whereas 
$\bar{R}^{2}(t)$ is averaged only over the surviving runs. In
accordance with the scaling ansatz for models with a unique
absorbing state \cite{torre,pgdp} it follows that these quantities
have the following scaling form,

\begin{eqnarray}
 P(t) & \propto & t^{-\delta } \Phi(\Delta t^{1/\nu_{\parallel}}),  \\
 \bar{n}(t) & \propto & t^{\eta }\Psi(\Delta t^{1/\nu_{\parallel}}), \\
 \bar{R}^{2}(t) & \propto & t^{z}\Theta(\Delta t^{1/\nu_{\parallel}}), 
\end{eqnarray}
 
where $\Delta = |p - p_{1}|$ is the distance from the critical point,
and $\nu_{\parallel}$ is the time-like correlation length exponent.
If the scaling functions $\Phi$, $\Psi$, and $\Theta$ are non-singular 
at the origin it follows that $P(t)$, $\bar{n}(t)$, and $\bar{R}^{2}(t)$
behave as power-laws at $p_{1}$ with critical exponents $-\delta$,
$\eta$, and $z$, respectively, for $t \rightarrow \infty$. Generally 
one has to expect corrections to a pure power law behaviour so that,
e.g., $P(t)$ is more accurately given by \cite{pgdp}
 
   \begin{equation}
      P(t) \:\propto \:t^{-\delta }(1 \ + \ at^{-1} \ 
     + \ bt^{-\delta '}\ + \ \cdots \ )
   \end{equation}
 
and similarly for $\bar{n}(t)$ and $\bar{R}^{2}(t)$. More 
precise estimates for the critical exponents can be obtained 
if one looks at local slopes
 
\begin{equation}
  -\delta (t) \:=\:\frac{\log[P(t)/P(t/m)]}{\log(m)},
  \label{eq:localslope}
\end{equation}
 
and similarly for $\eta (t)$ and $z(t)$. In a plot of the local 
slopes vs $1/t$ the critical exponents are given by the intercept 
of the curve for $p_{1}$ with the $y$-axis. The off-critical 
curves often have very notable curvature, i.e., one will see the
curves for $p < p_{1}$ veering downward while the curves 
for $p > p_{1}$ veer upward. This enables one to obtain 
accurate estimates for $p_{1}$ and the critical exponents.
In Fig.~1 I have plotted the local slopes for various values of
$p$. From the plot of $\eta (t)$ it is clear that the two lower
curves, corresponding to $p = 0.1781$, and 0.1782, veers downward
showing that $p_{1} > 0.1782$. Likevise the upper curve, $p=0.1785$,
has a pronounced upwards curvature. Though it is less evident it also
seems that the curve for $p = 0.1784$ veers upwards. All in all I 
conclude that $p_{1} = 0.1783(1)$. This estimate differs quite a
bit from that of Yaldram {\em et al.} ($p_{1}=0.185(2)$), which is
probably due to the slightly different algorithms. Note that
since NO adsorption is less efficient in my algorithm one would
expect my estimate for $p_{1}$ to be smaller, as is also 
observed in the simulations. From the intercept of the
critical curves with the $y$-axis I estimate $\delta = 0.45(1)$,
$\eta = 0.220(5)$ and $z = 1.12(1)$. These values agrees very well
with those obtained from computer simulations of directed 
percolation in (2+1)-dimensions \cite{pgdp}, $\delta = 0.460(6)$,
$\eta = 0.214(8)$ and $z = 1.134(4)$. 

From these results it seems reasonable to conclude that the CO-NO 
model belongs to the DP universality class. However, due to the 
somewhat arbitrary choice of the initial configuration employed
in the time-dependent simulations it would be nice to validate
this conclusion through other means. To this end I have also 
performed extensive steady-state simulations using a finite-size
scaling analysis. Finite-size scaling, 
though originally developed for equilibrium systems, is also 
applicable to nonequilibrium second-order phase transitions as 
demonstrated by Aukrust {\em et. al.} \cite{aukrust}. Their 
method was later applied to models with infinitely many 
absorbing states \cite{pcpprl,pcppre}. As in equilibrium 
second-order phase transitions one assumes that the (infinite-size)
nonequilibrium system features a length scale which diverges at 
criticality as, $\xi(p) \propto \Delta^{-\nu_{\perp}}$, 
where $\nu_{\perp}$ is the correlation length exponent in the space 
direction. The basic finite-size scaling ansatz is that the various 
quantities depend on system-size only through the scaled length
$L/\xi$, or equivalently through the variable 
$\Delta L^{1/\nu_{\perp}}$, where $L$ is the linear extension of 
the system. Thus we assume that the density of empty sites 
(which will be used as the order parameter of the model) 
depends on system size and distance from the critical point as:
	     
\begin{equation}
  \rho_{s}(p,L) \propto L^{-\beta/\nu_{\perp}} 
  {\cal F}(\Delta L^{1/\nu_{\perp}}), \label{eq:rhofss}
\end{equation}
such that at $p_{1}$
 \begin{equation}
  \rho_{s}(p_{1},L) \propto L^{-\beta/\nu_{\perp}}. \label{eq:rhocr}
\end{equation}

In $\rho_{s}$, and other quantities, the subscript $s$ indicates an 
average taken over the {\em surviving} samples. Fig.~2 shows a plot 
of the average concentration of particles $\log_{2}[\rho_{s}(p_{1},L)]$  
as a function of $\log_{2}L$ at the critical point, $p_{1}=0.1783$.
All simulations were performed on lattices of size $L\times L$ 
using periodic boundary conditions. The maximal number of timesteps 
in each trial, $t_{M}$, and number independent samples, $N_{S}$, varied 
from $t_{M}=300$, $N_{S}=50,000$ for $L=8$ to $t_{M} = 125,000$, 
$N_{S}=500$ for $L=256$. The slope of the line drawn in the figure 
is $\beta/\nu_{\perp} = 0.81$, which comes from the DP estimate 
$\beta/\nu_{\perp} = 0.81(2)$, using the earlier cited estimate for
$\beta$ and $\nu_{\perp} = 0.729(8)$ \cite{pgdp}. The data falls 
very nicely on the line drawn using the DP estimate thus confirming 
that the model belongs to the DP universality class.

Near the critical point the order parameter fluctuations grow like 
a power law,
$\chi_{s} = L^{d}(\langle \rho^{2} \rangle -
\langle \rho \rangle^{2}) \propto \Delta^{\gamma}$,
from which we expect the following finite-size scaling form,

\begin{equation}
  \chi_{s}(p,L) \propto L^{\gamma/\nu_{\perp}} 
  {\cal G}(\Delta L^{1/\nu_{\perp}}), \label{eq:chifss}
\end{equation}
such that at $p_{1}$
 \begin{equation}
  \chi_{s}(p_{1},L) \propto L^{\gamma/\nu_{\perp}}. \label{eq:chicr}
\end{equation}

Fig. 3 shows a plot of $\log_{2}[\chi_{s}(p_{1},L)]$ vs $\log_{2}L$.
The slope of the straigth line is 0.39 as obtained from the DP
value $\gamma/\nu_{\perp} = 0.39(2)$, where I used that
$\gamma = \gamma^{DP}-\nu_{\parallel} = 0.285(11)$ with 
$\gamma^{DP} = 1.571(6)$ \cite{pgdp}. The excellent agreement
between the data and the DP-expectation confirms the DP critical
behaviour of this model.

One expects a characteristic time for the system, say the 
relaxation time, to scale like

\begin{equation}
  \tau(p,L) \propto L^{-\nu_{\parallel}/\nu_{\perp}} 
  {\cal T}(\Delta L^{1/\nu_{\perp}}), 
\end{equation}
such that at $p_{1}$
 \begin{equation}
  \tau(p_{1},L) \propto L^{-\nu_{\parallel}/\nu_{\perp}}. 
\end{equation}

In Fig. 4 I have plotted $\log_{2}[\tau_{h} (p_{1},L)]$, where
$\tau_{h}$ is the time it takes for half the samples to enter an 
absorbing state, as a function of $\log_{2} L$. The slope of the line drawn in the figure is $\nu_{\parallel}/\nu_{\perp} = 1.764$, as
obtained from the DP estimate \cite{pgdp} 
$\nu_{\parallel}/\nu_{\perp} = 1.764(7)$. The DP estimate is derived
from the scaling relation $\nu_{\parallel}/\nu_{\perp} = 2/z$ using
the earlier cited estimate for $z$. As can be seen the data
for the CO-NO model is again fully compatible with DP critical
behaviour.
 
One may also study the dynamical behaviour by looking at the time 
dependence of $\rho_{s}(p_{1},L,t)$. For $t \gg 1$ and $L \gg 1$ 
one can assume a scaling form

\begin{equation}
  \rho_{s}(p_{1},L,t) \propto 
  L^{-\beta/\nu_{\perp}}{\cal H}(t/L^{\nu_{\parallel}/\nu_{\perp}}). 
  \label{eq:rhofssall}
\end{equation}

At $p_{1}$ the system shows a power law behaviour for 
$t < L^{\nu_{\parallel}/\nu_{\perp}}$ before finite-size effects 
become important. Thus for $L \gg 1$ and 
$t < L^{\nu_{\parallel}/\nu_{\perp}}$, 
$\rho(p_{1},L,t) \propto t^{-\theta}$. From Eq.~(\ref{eq:rhofssall}) 
we see that this is the case for large $L$ only if 
$\theta  = \beta/\nu_{\parallel}$. It can be shown \cite{torre}
that this ratio also equals the critical exponent $\delta$.
Fig.~5 shows the short-time evolution of the concentration of 
empty sites at $p_{1}$ with $L=256$, $t_{M}=10,000$, and 
$N_{S} = 1000$. The asymptotic behaviour is consistent with
$\theta = 0.45$, as seen from the slope of the line. This
estimate agrees well with the value for directed percolation
$\theta = \delta = 0.460(6)$, or the estimate $\delta=0.45(1)$
obtained from the time-dependent simulations presented above. 

In conclusion, we have provided very convincing evidence that
the critical exponents of the two dimensional CO-NO model are the 
same as those of directed percolation in (2+1)-dimensions. This is
the first time that a two dimensional multi-component model
with infinitely many absorbing states has been firmly placed in the
DP universality class. This results lends further support to the 
extensions of the DP conjecture to models with multiple components
\cite{zgbrg} and/or infinitely many absorbing states
\cite{pcpprl,pcppre}, at least in cases where the absorbing states 
can be characterized by the vanishing of a unique quantity.

\newpage

\begin{center} {\Large \bf Figure Captions} \end{center}

{\bf Figure 1} Local slopes $-\delta (t)$ (upper panel), $\eta (t)$ 
(middle panel), and $z(t)$ (lower panel), as defined in 
Eq.~\protect{\ref{eq:localslope}} with $m=5$. Each 
panel contains five curves with, from bottom to top, $p=0.1781$, 
0.1782, 0.1783, 0.1784 and 0.1785.

{\bf Figure 2} The concentration of empty sites $\log_{2}[\rho_{s}(p_{1},L)]$ {\em vs} $\log_{2} L$. The slope
of the straight line is $\beta/\nu_{\perp} = 0.81$.

{\bf Figure 3}  The fluctuations in the concentration of 
empty sites $\log_{2}[\chi_{s}(p_{1},L)]$ {\em vs} $\log_{2}L$.
The slope of the straight line is $\gamma/\nu_{\perp} = 0.39$.

{\bf Figure 4}  The time before {\em half} the samples
enter an absorbing state $\log_{2}[\tau_{h}(p_{1},L)]$ 
{\em vs} $\log_{2}L$. The slope of the straight line is 
$\nu_{\parallel}/\nu_{\perp} = 1.764$.
 
{\bf Figure 5} Log-log plot of $\rho_{s}(p,L,t)$, for 
$p=p_{1}=0.1783$ and $L=256$, as a function of $t$. 
The slope of the straight line is $\theta = 0.45$.


\newpage
\begin{thebibliography}{99}

\bibitem{dpbook} Several articles about directed percolation may 
be found in {\em Percolation Structures and Processes}, edited by
G. Deutsher, R. Zallen, and J.Adler, Annals of the Israel
Physical Society Vol.~5 (Hilger, Bristol, 1983).

\bibitem{blease} J. Blease, J. Phys. C {\bf 10}, 917, 923, 
and 3461 (1977).

\bibitem{cardy} J. L. Cardy and R. L. Sugar, J. Phys. A \bf13\rm, 
L423 (1980).

\bibitem{kinzel} W. Kinzel, Z. Phys. B {\bf 58}, 229 (1985).

\bibitem{gribov} V. N. Gribov, Sov. Phys. JETP {\bf 26}, 414 (1968);
 V. N. Gribov and A. A. Migdal, Sov. Phys. JETP {\bf 28}, 784 (1969).

\bibitem{brower} R. C. Brower, M. A. Furman, and M. Moshe,
Phys. Lett. B {\bf 76}, 213 (1978).

\bibitem{harris} T. E. Harris, Ann. Prob. {\bf 2}, 969 (1974).

\bibitem{liggett} T. M. Liggett, {\em Interacting Particle
Systems} (Springer-Verlag, New York, 1985).

\bibitem{durrett} R. Durrett, {\em Lecture Notes on Par\-ticle
Sy\-stems and Per\-co\-la\-tion} (Wadsworth Pub. Co.,
Pacific Grove, CA, 1988).

\bibitem{schlogl} F. Schl\"{o}gl, Z. Phys. B \bf253\rm, 147 (1972).

\bibitem{torre} P. Grassberger and A. de la Torre, Ann. Phys. (NY)
 \bf 122\rm, 373 (1979).

\bibitem{janssen81} H. K. Janssen, Z. Phys. B \bf42\rm, 151 (1981).

\bibitem{pgrg} P. Grassberger, Z. Phys. B \bf47\rm, 365 (1982).

\bibitem{zgb} R. M. Ziff, E. Gulari, and Y. Barshad,
Phys. Rev. Lett. \bf56\rm, 2553 (1986).

\bibitem{bbc} R. Bidaux, N. Boccara, and H. Chat\'{e}, 
Phys. Rev. A \bf39\rm, 3094 (1989). \\
I. Jensen, Phys. Rev. A {\bf43}, 3187 (1991).

\bibitem{rondif} R. Dickman, Phys. Rev. B {\bf40}, 7005 (1989).

\bibitem{aukrust} T. Aukrust, D. A. Browne, and I. Webman,
Phys. Rev. A {\bf 41}, 5294 (1990)

\bibitem{tripann} R. Dickman, Phys. Rev. A {\bf42}, 6985 (1990).

\bibitem{tripcrea} R. Dickman and Tania Tom\'{e},
Phys. Rev. A {\bf 44}, 4833 (1991).

\bibitem{baw} H. Takayasu and A. Yu Tretyakov, Phys. Rev. Lett.
{\bf 68} 3060 (1992). \\
I. Jensen, Phys. Rev E {\bf 47}, R1 (1993). \\
I. Jensen, J. Phys. A {\bf 26}, 3921 (1993).

\bibitem{dollarsdimes} H. Park, J. K\"{o}hler, I-M Kim, 
D. ben-Avraham, and S. Redner, J. Phys. A {\bf 26}, 2071 (1993).

\bibitem{zhuo} J. Zhuo, S. Redner, and H. Park,
J. Phys. A {\bf 26}, 4197 (1993).

\bibitem{dimertrimer} J. K\"{o}hler and D. ben-Avraham,
J. Phys. A {\bf 24}, L621 (1991);
D. ben-Avraham and J. K\"{o}hler, J. Stat. Phys. {\bf 65}, 839 (1992).

\bibitem{dimerdimer} E. V. Albano, J. Phys. A {\bf 25}, 2557 (1992);
A. Maltz and E. V. Albano, Surf. Science {\bf 277}, 414 (1992);
E. V. Albano, J. Stat. Phys. {\bf 69}, 643 (1992).

\bibitem{pcpprl} I. Jensen, Phys. Rev. Lett. {\bf 70}, 1465 (1993). 

\bibitem{pcppre} I. Jensen and R. Dickman, Phys. Rev. E {\bf 48}, 1710 (1993).

\bibitem{cono} K. Yaldram, K. M. Khan, N. Ahmed, and M. A. Kkan,
J. Phys. A {\bf 26}, L801 (1993).

\bibitem{pgdp} P. Grassberger, J. Phys. A \bf22\rm, 3673 (1989).

\bibitem{mulscal} J. F. F. Mendes, R. Dickman, M. Henkel, and
M. C. Marques, {\em Generalized Scaling for Models with Multiple
Absorbing States}, preprint 1993.

\bibitem{zgbrg} G. Grinstein, Z. W. Lai, and D. Browne,
Phys. Rev. A {\bf 40}, 4820 (1989).

\end{thebibliography}
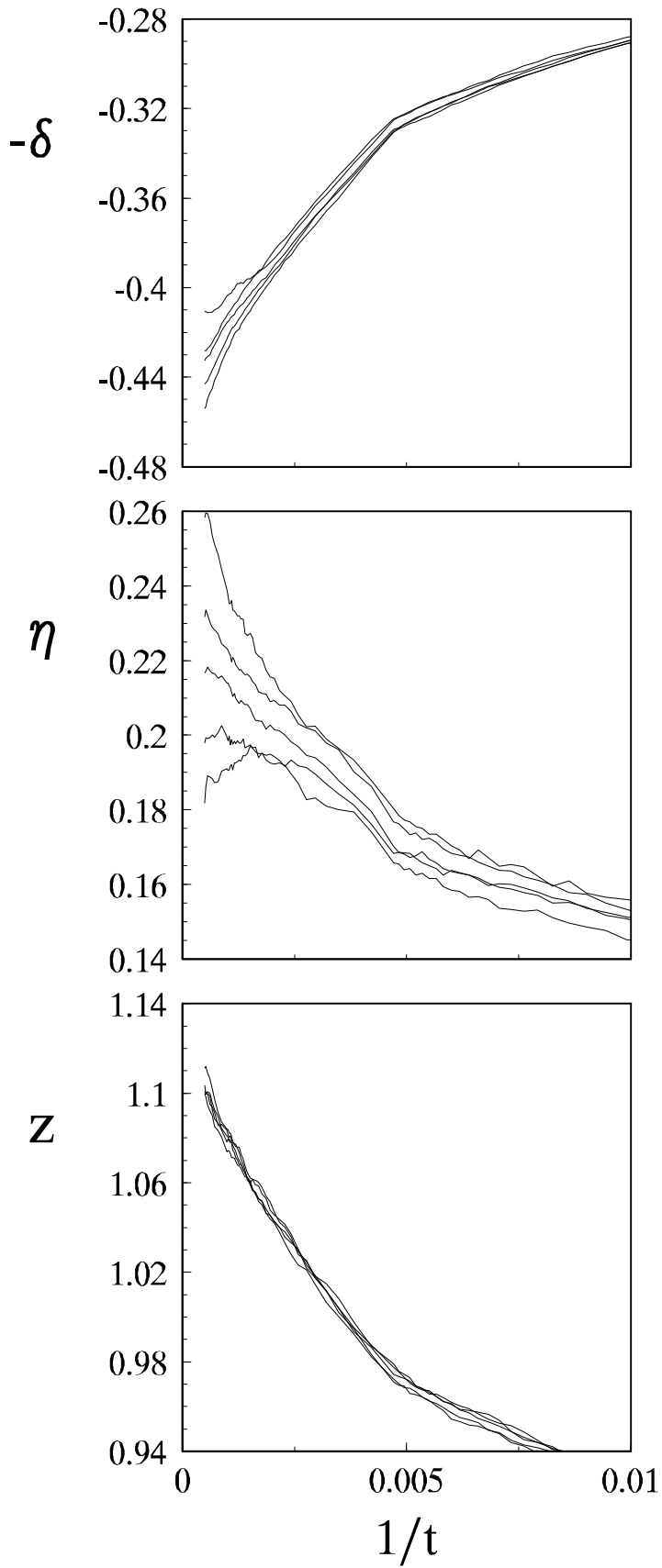
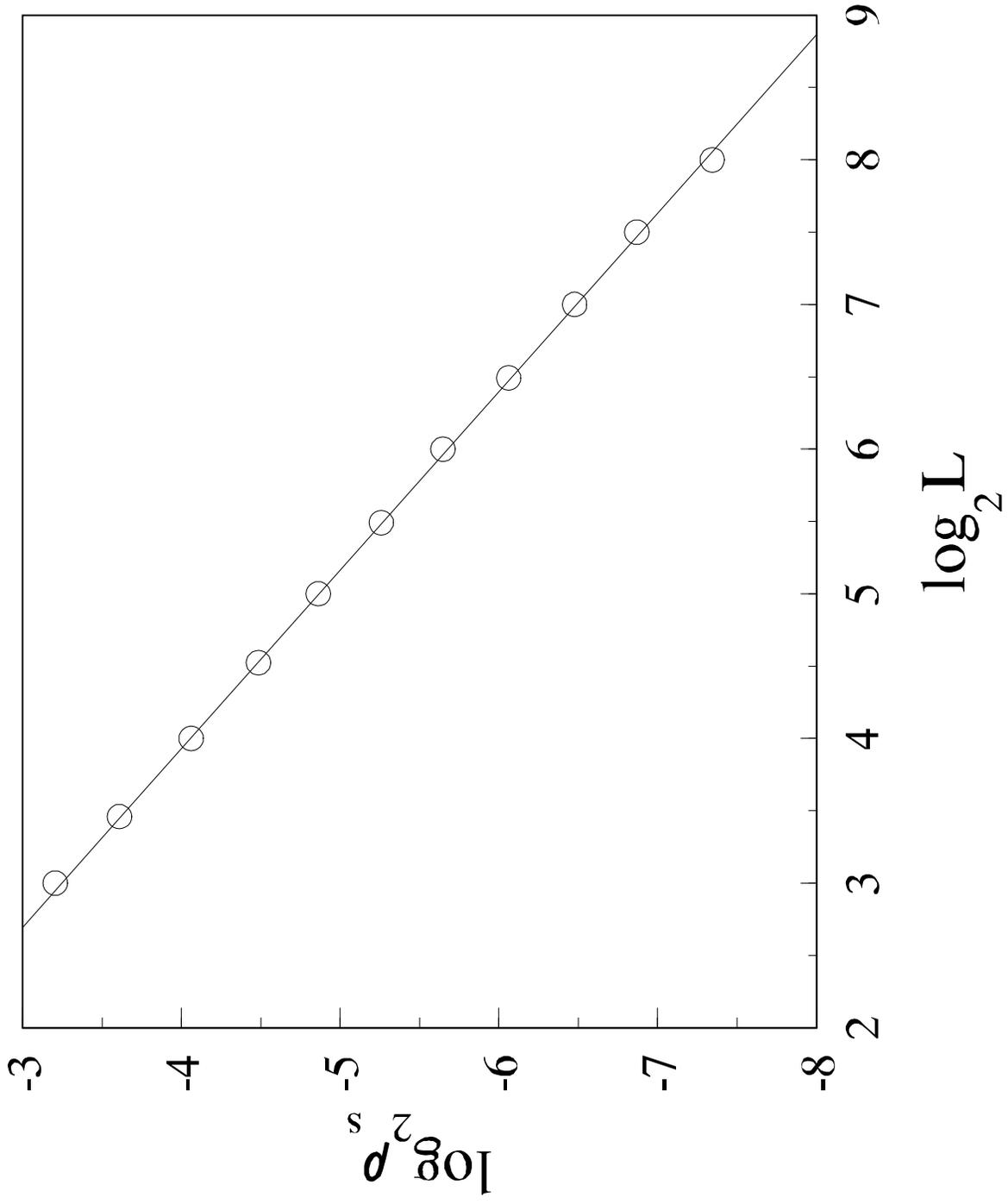
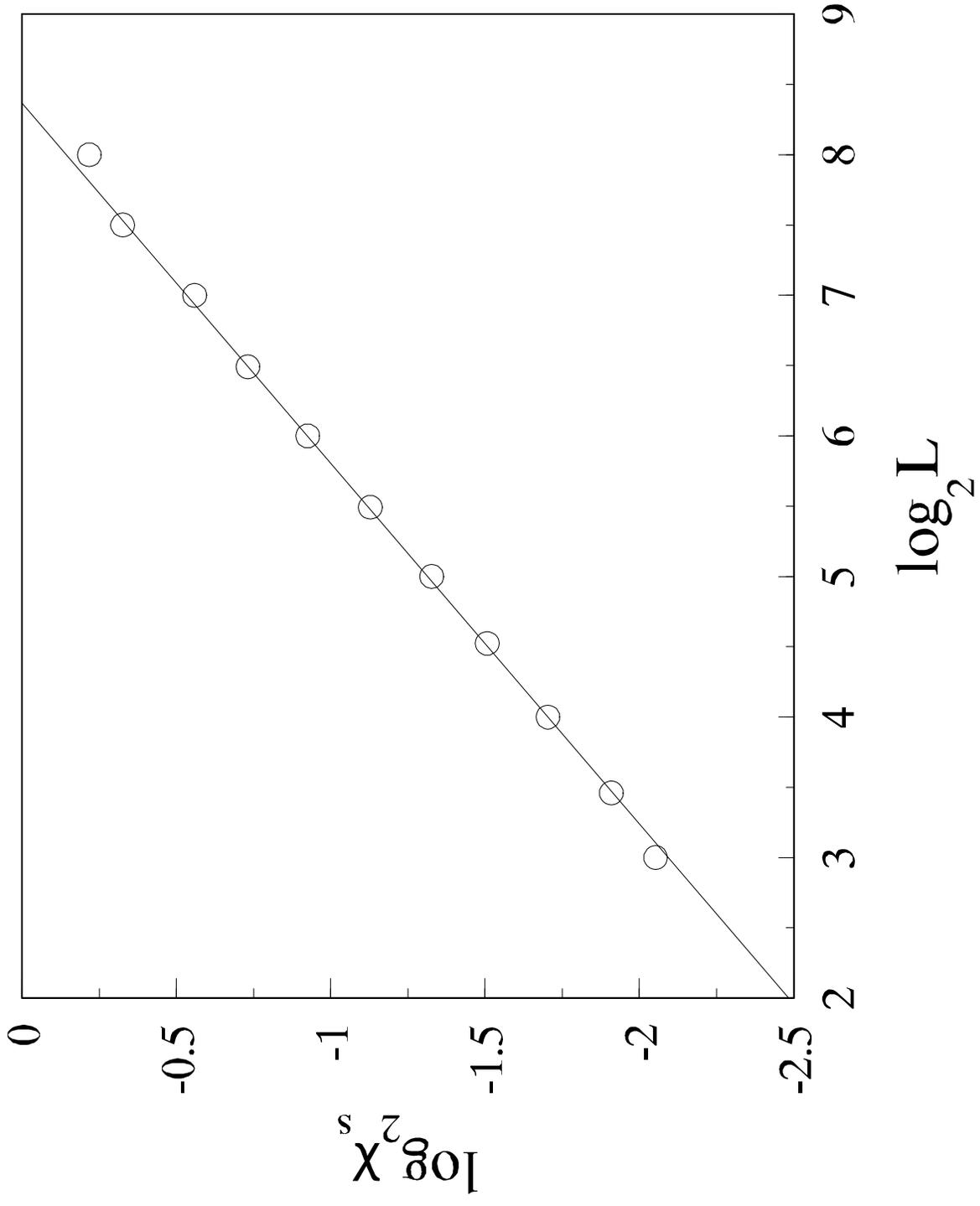
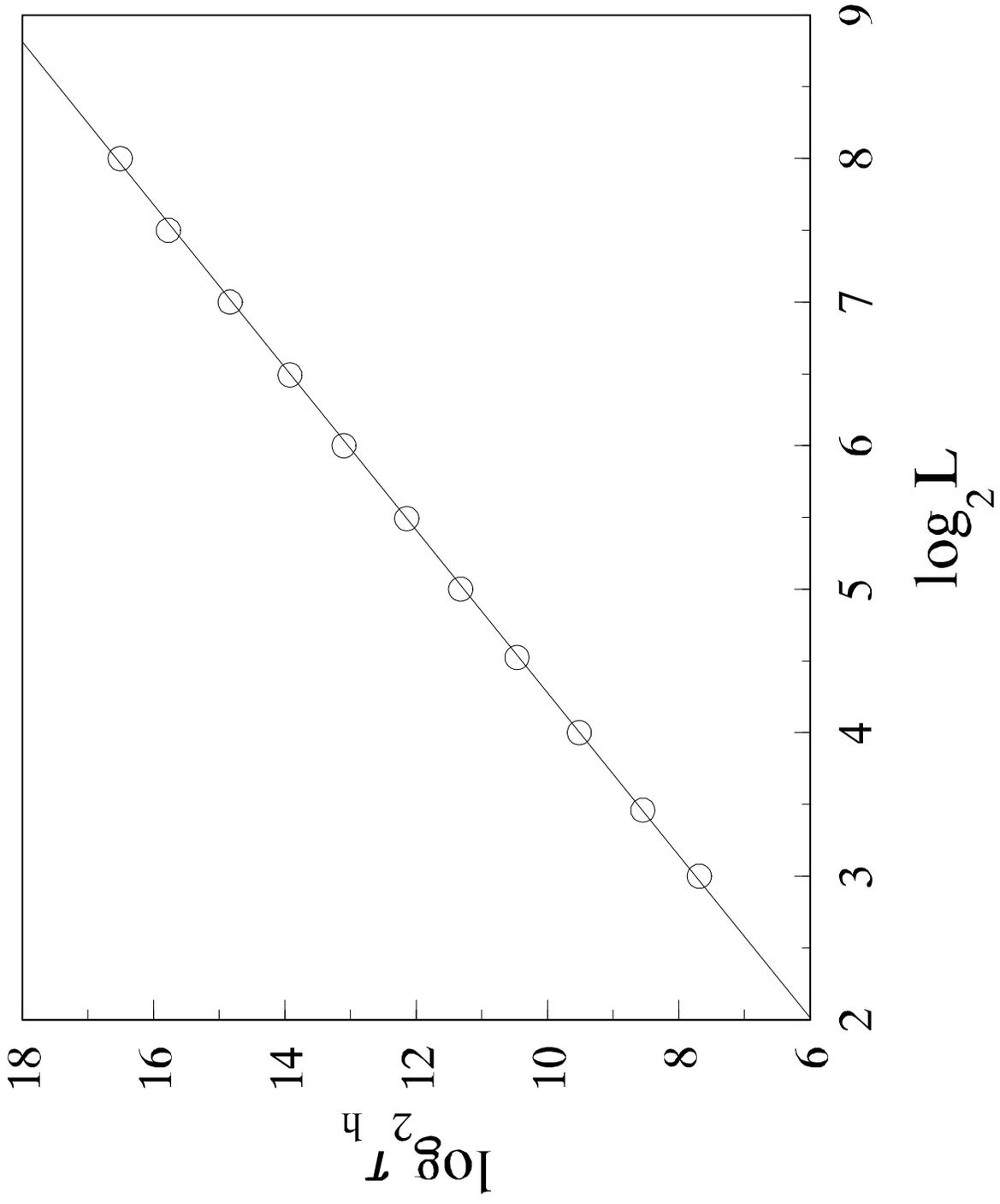
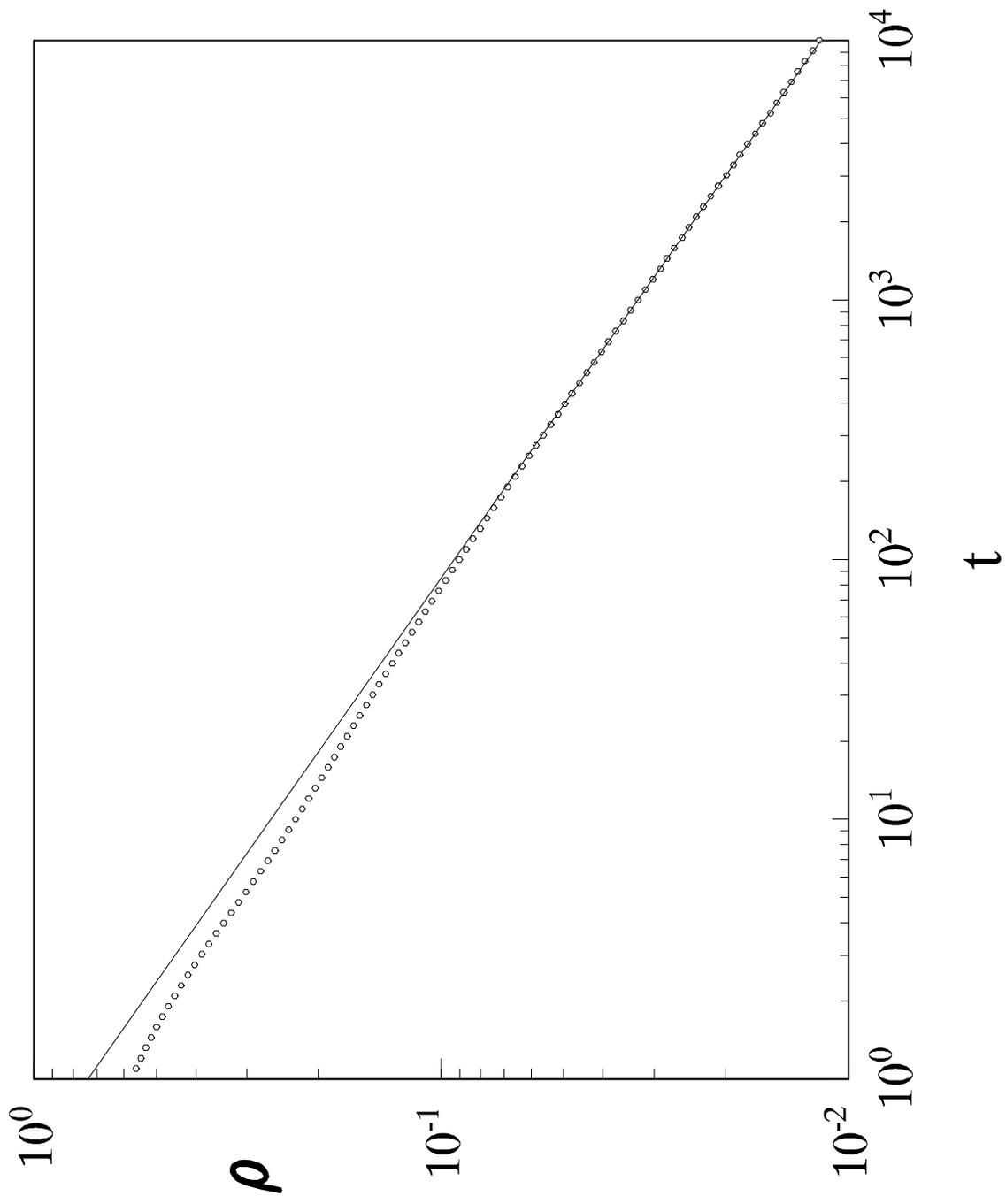
\end{document}